\documentclass[aps,pra,twocolumn,groupedaddress,floatfix]{revtex4-1}
\usepackage{graphicx}% Include figure files
\usepackage{dcolumn}% Align table columns on decimal point
\usepackage{bm}% bold math
\usepackage{amsmath}
\usepackage{color}
\newcommand{\sign}{\operatorname{sign}}
\DeclareMathOperator{\Tr}{Tr}
\begin{document}
%chiral effect in vesalago lens focusing: armchair vs zigzag graphene p-n junction
\title{ Valley-dependent Lorentz force and Aharonov-Bohm phase in strained graphene p-n junction }
\author{Sanjay Prabhakar,$^{1,2}$ Rabindra Nepal,$^1$ Roderick Melnik,$^{2,3}$ and  Alexey A. Kovalev$^1$ }
\affiliation{
$^1$Department of Physics and Astronomy, and Nebraska Center for Materials and Nanoscience,
University of Nebraska, Lincoln, Nebraska 68588, USA\\
$^2$The MS2Discovery Interdisciplinary Research Institute, M2NeT Laboratory, Wilfrid Laurier University, Waterloo, ON N2L 3C5, Canada\\
$^3$BCAM, Alameda Mazarredo 14, 48080 Bilbao, Spain
}
\date{January 27, 2019}

\begin{abstract}
Veselago lens focusing in graphene p-n junction is promising for realizations of new generation electron optics devices. However, the effect of the strain-induced Aharonov-Bohm interference in a p-n junction has not been discussed before. We provide an experimentally feasible setup based on the Veselago lens in which the presence of strain can result in both the valley-dependent Lorentz force and Aharonov-Bohm interference. In particular, by employing the Green's function and tight binding methods, we study the strain induced by dislocations and line defects in a p-n junction and show how the resulting Aharonov-Bohm phase and interference can be detected.
Furthermore, for a different strain configuration, e.g. corresponding to corrugated graphene, we find strong signatures of valley splitting induced by the fictitious magnetic field. Our proposal can be useful for mapping elastic deformations and defects, and for studying valley dependent effects in graphene.
\end{abstract}

% insert suggested PACS numbers in braces on next line
%\pacs{71.70.Ej, 73.61.Ey, 85.75.Hh}
% insert suggested keywords - APS authors don't need to do this
%\keywords{}

%\maketitle must follow title, authors, abstract, \pacs, and \keywords

\maketitle

\section{Introduction}
Two dimensional materials, like graphene and several others, can lead to realizations of optoelectronic devices operating at much higher frequencies compared to conventional devices~\cite{geim-nature07,coleman-science11}. A lot of theoretical and experimental research efforts have concentrated on graphene as it exhibits the half integer quantum Hall effect, non-zero Berry curvature, high mobility charge carriers (100 times higher than in Silicon), and other unique properties~\cite{novoselov05a,novoselov05,novoselov04,savage-nature12}.
It has been shown that CMOS devices made out of graphene are superior compared to the best silicon
devices of the same size~\cite{savage-nature12,novoselov04,liao-nature10}. The lack of bandgap, as conduction and valence bands touch each other at the Dirac point, makes graphene implausible for device applications. Nevertheless, by using several state-of-the-art engineering techniques, one can easily open small bandgaps. For example, bandgap opening  is achieved by considering the effect of spin-orbit coupling, or ripples and strain. Spintronics devices made from graphene nanoribbons possess a larger band gap opening at $\Gamma$-point~\cite{han-prl07,zhou-nature07,xia-nanoletters10,chen-nature15,ugeda-nature14,brey06}.

Graphene can be also used for realizations of electron optics devices, e.g., the transmission electron microscope. Here, a fine focusing of classical electron-hole trajectories can be achieved by making devices out of graphene p-n junctions~\cite{Lee.Park.eaNP2015,zhang18,cheianov07,jiang-nature17,tetienne-scienceAd17,betancur-prb17}.   The electric field control of electron-hole charge carriers in a transparent graphene p-n junction can utilize the idea of optical refraction at interfaces, where graphene acts as a material that  possesses properties of metamaterials with  negative refractive index~\cite{cheianov07,choi14,reijnders17,reijnders17a}. The negative refractive index arises in graphene p-n junction because the group velocity of electrons in the conduction band is opposite in direction to that of  holes in the valence band.  

The non-vanishing strain in graphene can induce fictitious vector potentials and gauge fields~\cite{Vozmediano.Katsnelson.eaPR2010,Guinea.Katsnelson.eaNP2010,Low.GuineaNL2010,juan13} and it can be utilized to measure the Aharonov-Bohm (AB) interference~\cite{Mesaros.Sadri.eaPRB2009,juan11}. Quantum interference phenomena can be revealed in mesoscopic conductivity measurements in a variety of setups~\cite{Aronov.Lyanda-GellerPRL1993,Bardarson.Brouwer.eaPRL2010,Frustaglia.RichterPRB2004,Molnar.Peeters.eaPRB2004,Kovalev.Borunda.eaPRB2007,Lin.Wang.eaPRB2017}. The effect of the Pancharatnam-Berry phase on the Veselago lens focusing in the armchair and zigzag graphene nanoribbons has been studied theoretically~\cite{choi14}. A possibility of spatial valley separation in electron-hole beam focusing in strained graphene p-n junctions has been suggested~\cite{tian17}.

%%%%%%%%%%%%%%%%%%%%%%%%%%%%%%%%%%%
\begin{figure*}
\includegraphics[width=12.0cm,height=8cm]{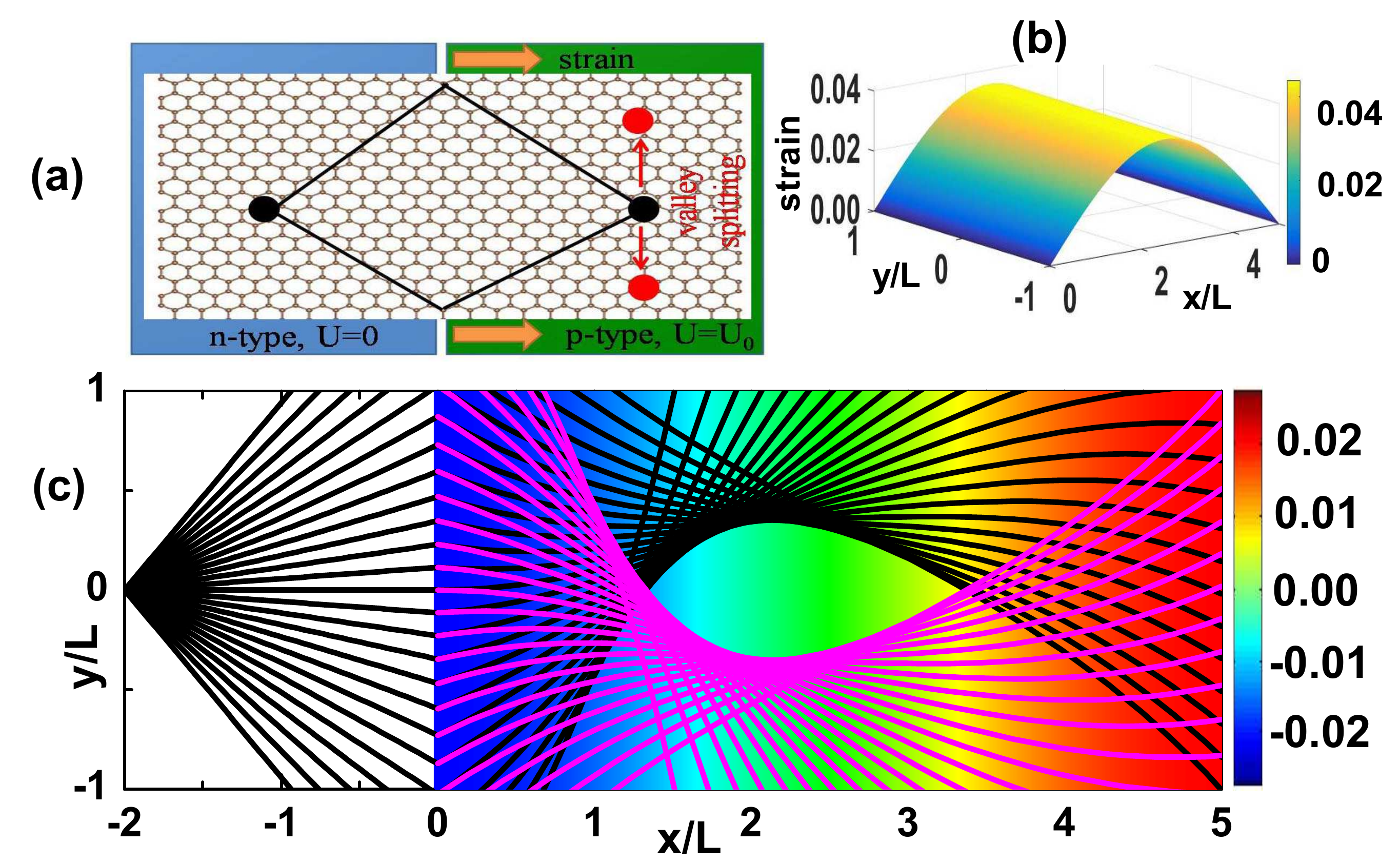}
\caption{\label{fig3} (a) A schematic setup for inducing valley splitting  in Veselago lens via corrugated strain along x-direction. (b) The distribution of strain $\varepsilon_{xx} = -Aq \sin(qx)$ in p-region. (c) Simulations of electron-hole beams trajectories in the presence of fictitious magnetic field. The black lines show the trajectories of holes for $K'$ valley while magenta lines for $K$ valley.  The dimensionless parameters are chosen as E=2$U_0$ with $U_0=2E$ with $E=2$, $A=-0.07$, $q=0.63$.
The values of dimensionless fictitious magnetic field are encoded in the background colors, and are in the range given by $B_s a L/\beta \phi_0$=(min,max)=(-0.028,0.028) where  $B_s=8.1$T for $L=100$nm.
}
\end{figure*}

%%%%%%%%%%%%

There is a variety of ways to control  strain in graphene. Both in-plane and out-of-plane components of strain tensor in graphene can be controlled in a desired fashion by applying in-plane and  out-of-plane deformations~\cite{bao-nature09,prabhakar16,christensen-prb15,prabhakar14} or by creating dislocations~\cite{Yazyev.LouiePRB2010,bonilla12science,carpio08njp,shallcross17nc} or line defects~\cite{Berger.RatschPRB2016,alexandre17prb,ishikawa18nc}.
A substrate (e.g. SiC) can induce a large strain due to lattice mismatch between graphene and the substrate~\cite{ni08,bastos16}.
Furthermore, applying compressive tensile edge stress through the armchair and zigzag boundaries  can also lead to the formation of ripples and wrinkles~\cite{bronsgeest15,cerda03,ryan17,prabhakar14,prabhakar16}.
Dangling bond sites at the edge of graphene can lead to the formation of edge strain due to adsorbtion of different organic materials~\cite{deepika15,lim15}.
Uniaxial strain in graphene can be induced by bending the substrate on which graphene is grown~\cite{huang09}.
Biaxial, localized strain can be induced by the atomic force microscope or scanning tunneling microscope tips~\cite{lee08,khestanova16}.
Tunable biaxial tensile and compressive strain can also be  induced by growing graphene  on a piezoelectric substrate and by controlling the bias voltage~\cite{ding10}.
Finally, tensile or compressive biaxial strain in graphene can be induced by employing the thermal expansion coeficient mismatch between the graphene and the substrate (e.g. SiC)~\cite{ferralis08,boyd15}.

In this paper, using the Green's function and tight binding methods, we show that Veselago lens can be used for mapping strain, e.g., produced by in-plane ripples, line defects and dislocations~\cite{meng13,levy10,juan11,verbiest16,Yazyev.LouiePRB2010,bonilla12science,carpio08njp,shallcross17nc,Berger.RatschPRB2016,alexandre17prb,ishikawa18nc}.  
The presence of strain leads to a fictitious vector potential which in turn can lead to a Lorentz force and to accumulation of the Berry phase. Below, we show that both effects can be separately identified in a graphene p-n junction. 
To this end, we first study the valley separation and signatures of strong Lorentz force in the trajectories of graphene holes and electrons subjected to strain, e.g. in corrugated graphene~\cite{meng13}, which could have implications for the field of valleytronics. Next, we study the Berry phase accumulation in a setup containing line defects and dislocations and provide an experimental setup for AB phase measurement by employing the Veselago lens focusing.

The paper is organized as follows. In Sec.~\ref{theoretical-model}, we provide a detailed theoretical formulation of the Green's function approach applicable to a strained graphene p-n junction. We then study the effect of strain  on the diffraction patterns  of charge carriers in such graphene p-n junctions. First, we show how strain engineering can lead to valley splitting. The Green's function approach is further used to describe the AB phase for strain induced by line defects or dislocations. We also perform tight binding simulations to confirm our predictions numerically. In Sec.~\ref{conclusion}, we give our conclusions.

%%%%%%%%%%%%%%%%%%%%%%%%%%%%%%%%%%%
\begin{figure}
\includegraphics[scale=0.7]{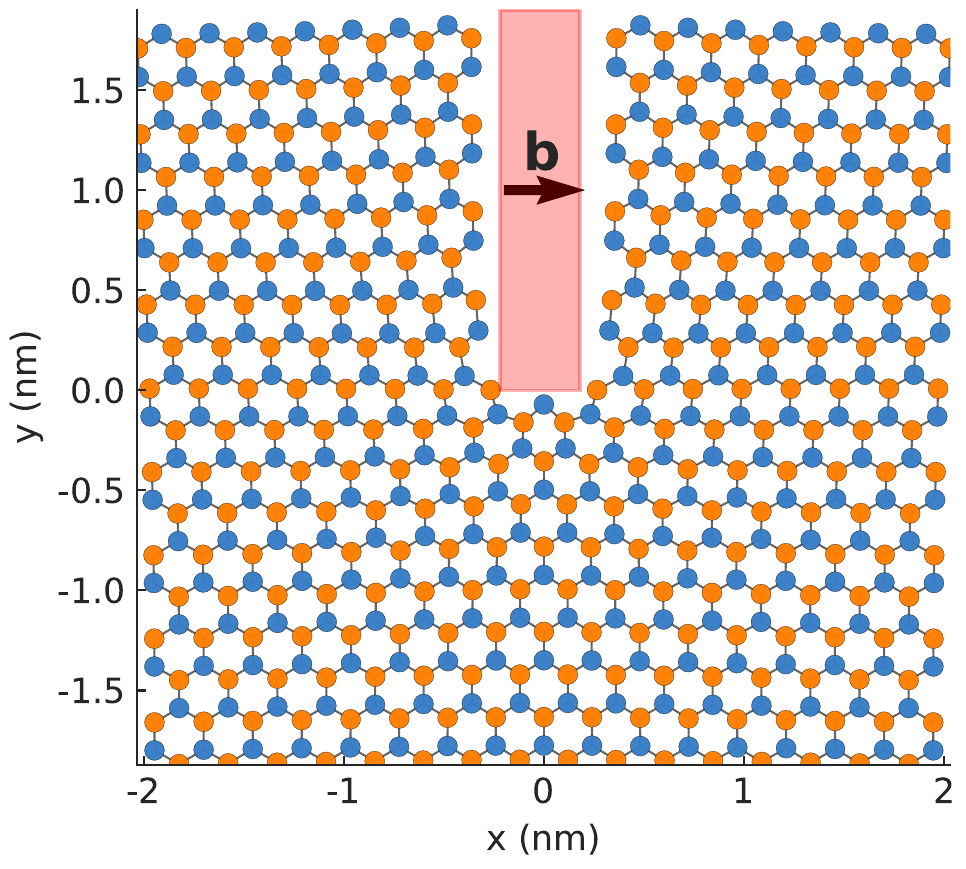}\label{strain1}
\caption{\label{strain1} Schematics of a graphene sheet with a p-n junction at $x=0$. The strain is induced by a graphene insertion shown as a filled rectangle. Dislocations in graphene result in insertions (or cuts) described by the Burgers vector, e.g., $\vec{b}=n \vec{a}_1 +m \vec{a}_2$, where
$\vec{a}_1$ and $\vec{a}_2$ are translation vectors of graphene lattice. A chemically induced line defect can also result in the deformation shown in this figure.
}
\end{figure}

%%%%%%%%%%%%

\section{Strained p-n junction}\label{theoretical-model}
%%%Vessalago lens

%\noindent
%\textit{\underline{Theoretical Model}.}
In this section, we study a strained graphene p-n junction and try to identify signatures of the fictitious Lorentz force (see Fig.~\ref{fig3}) and Berry phase (see Figs.~\ref{strain1} and \ref{ABtb}). In the continuum limit, after expanding the momentum close to the $K(K')$ point in the Brillouin zone, the  Hamiltonian  for $\pi$ electrons at the $K(K')$ point in strained graphene sheet with zigzag edge reads  as~\cite{suzuura-prb02}:
%Equation 38 from nato09 reference
\begin{equation}
H= v_F\left(\sigma_x P_x +\tau \sigma_y P_y\right)+U(x), \label{H}
\end{equation}
where $\mathbf{P}=\mathbf{p}-e \mathbf{A}$ with $\mathbf{p}=-i\hbar \mathbf{\nabla}$ being the canonical momentum operator and $\mathbf{A}=\beta \phi_0 \left(-2\varepsilon_{xy},\varepsilon_{yy}-\varepsilon_{xx},0\right)/a$ is the vector potential induced by  strain tensor, $\phi_0=2 \pi \hbar/e$ is the fundamental unit of flux, $\varepsilon_{ij}=1/2[\partial_ju_i+\partial_iu_j+(\partial_i h)(\partial_j h)]$ is strain tensor expressed in terms of in-plane and out-of-plane displacements, $\mathbf{u}$ and $h$, and $\tau=\pm 1$ for the $K(K')$ valley \cite{suzuura-prb02,stegmann-njp16}. Here $U(x)=0$ for $x<0$ and $U(x)=U_0$ for $x>0$,  $a$ is the lattice constant, $\beta=-\partial \ln t/\partial \ln a \approx 2$ describes the change in the hopping amplitude as the bond length changes and  $t$ is  the nearest neighbor hopping parameter.
%%1st-referee-2nd comment
%%1st-referee-2nd comment
%%1st-referee-2nd comment
%%1st-referee-2nd comment

In this paper, we limit our consideration to pure in-plane deformations; however, a case with more general strain should lead to similar physics. To get a clear signature of the Lorentz force, we consider strain along x-direction, as shown in Fig.~\ref{fig3}, (i.e., only $u_x$ is  non-vanishing) which leads to the valley splitting. Experimentally, such strain can be realized in an armchair corrugated graphene nanoribbon~\cite{meng13}. To get clear signatures of the Berry phase, we consider strain induced by dislocations or line defects, as shown in Fig.~\ref{strain1}, which results in the AB-like interference effects.

Throughout the paper, we use dimensionless parameters as follows: $\tilde{x}=x/L$, $\tilde{y}=y/L$, $\tilde{x_s}=x_s/L$, $\tilde{k_y}=k_y L$, $\tilde{E}=EL\iota/\hbar v_F$ with $\iota=\hbar v_F/E_0L$, $\tilde{U}=U_0L\iota/\hbar v_F$ and  $\tilde{\Psi}=E_0L^2\Psi$. Here $L$ is the width of the graphene nanoribbon and $E_0$ is the typical energy scale of the problem.

\subsection{Valley splitting due to Lorentz force}
To identify the effect of strain, we consider the eigenvalue problem, $H \Psi=E \Psi$, where the spinor wavefunction  of Hamiltonian~(\ref{H}) can be written as,
%\begin{eqnarray}
$
\Psi\left(r\right)=\exp{\left(ik_y y\right)} \left(\begin{array}{c}
 \Phi_A\left(x\right)~
 \Phi_B\left(x\right)
\end{array}\right)^T.
$
Thus  from (\ref{H}),  we write two coupled equations  as
\begin{eqnarray}
-i\hbar v_F \left( \partial_x +\tau k_y+\beta\varepsilon_{xx}/a\right)\Phi_B=\left(E-U_0\right) \Phi_A, \label{hbar-vF-1} \\
-i\hbar v_F \left(\partial_x-\tau k_y-\beta\varepsilon_{xx}/a\right)\Phi_A=\left(E-U_0\right) \Phi_B. \label{hbar-vF-2}
\end{eqnarray}
%%%%%%%%%%%%%%%%%%%%%%%%%%%%%%%%%%%%%%%%%%%%
\begin{figure*}
\includegraphics[width=15.0cm,height=9.5cm]{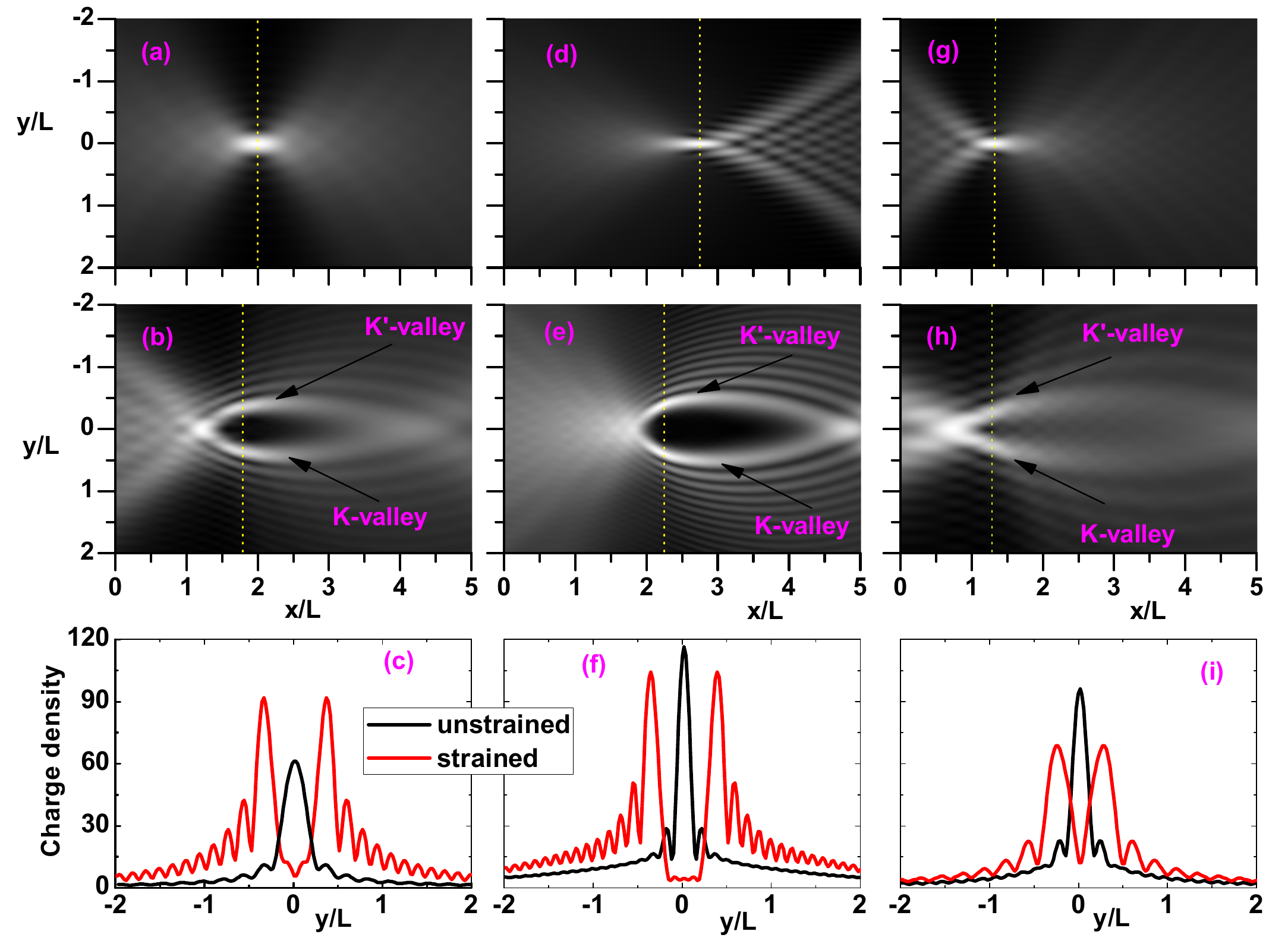}
\caption{\label{fig4Diff} Diffraction patterns of particle density for unstrained in (a,d,g) and  valley separation due to applied strain in (b,e,h) near hole focal point in graphene p-n junction. The cross section  plot in (c,f,i) along y-direction at the dotted lines captures the maxima of the particle density. Here we have chosen the dimensionless parameters, $A=-0.07$, $q=0.63$, $\iota=0.0639$, $U_0=2E$ with $E=2$ in (a,b,c), $E=2$, $U_0=4.5$ in (d,e,f) and $E=2$, $U_0=3.5$ in (g,h,i). For graphene p-n junction, these numbers correspond to $E=203$ meV and $L=100$ nm.
}
\end{figure*}
%%%%%%%%%%%
%%%%%%%%%%%%%%%%%%%%%%%%%%%%%%%%%%%%%%%%%%%
Now, we apply operator
%\begin{equation}
%$i\hbar v_F \left(-\partial_x +k_y +\beta\varepsilon_{xx}/a  \right)$
%\end{equation}
%from left in (\ref{hbar-vF-1})  and operator
%\begin{equation}
$-i\hbar v_F \left(\partial_x +k_y +\beta\varepsilon_{xx}/a  \right)$
%\end{equation}
from left in (\ref{hbar-vF-2}) and write a single decoupled second order partial differential equation  as:
\begin{equation}
\partial_x^2 \Phi_{A}=-\left[\left(\frac{E-U_0}{\hbar v_F}\right)^2-k_y^2-e_{xx}- \frac{\beta}{a}\chi(x)\right]\Phi_{A}, \label{PhiA}
\end{equation}
%%%%%%%%%%%%%%%%%%%
where
$
e_{xx} = \left(\beta\varepsilon_{xx}/a\right)^2+\tau 2\beta \varepsilon_{xx}k_y/a,
$
and $\chi (x) = -Aq^2\cos(qx).$ Also $\varepsilon_{xx}=\partial_x u_x$, where $u_x=A\cos(qx)$ with $A$ being the amplitude of the ripple wave and $q=2\pi/\lambda$, where $\lambda$ is the wavelength of the ripple wave.
Here the non-vanishing strain induces fictitious magnetic  fields~\cite{suzuura-prb02,meng13,prabhakar16}.
When the fictitious magnetic fields are comparable to $50\,$T, they  induce Landau levels~\cite{levy10,meng13}. In the opposite limit of weak fictitious magnetic fields considered here, one can write the solutions of Eq.~(\ref{PhiA}) in terms of semiclassical trajectories~\cite{juan11}. We introduce source term, $J(x)= (\alpha_1~\alpha_2)^T$ $\delta(x-x_s)$, in ~(\ref{H}) and write its solution  in terms of Green's functions~\cite{reijnders17},  $\Psi(x)=G(x,x_s)~(\alpha_1~\alpha_2)^T $,
where $\alpha_1$ and  $\alpha_2$ are constants and
\begin{eqnarray}\nonumber
G(x,x_s)=\frac{i}{4\pi \iota^2}\int_{-k_m}^{k_m} && dk_y  \left(\begin{array}{cc}
 e^{i(\phi-\theta)/2} &  e^{-i(\phi+\theta)/2} \\
 e^{i(\phi+\theta)/2} &  e^{-i(\phi-\theta)/2}
\end{array}\right) \\
 &&\times \frac{1}{\cos((\phi+\theta)/2)}} e^{iS(k_y,x,y)/\iota.\label{G}
\end{eqnarray}
Here $k_m$ is the maximum value of $k_y$,   $\phi$ and $\theta$ are angles made by incident electrons  and transmitted holes  at the interface and $\iota=\hbar v_F/E_0L$ is a constant. The classical action, $S(k_y,x,y)$, is written as
\begin{equation}
  S(k_y,x,y) = -x_s\sqrt{\left(\frac{E}{\hbar v_F}\right)^2-k_y^2}-\int^x p_h(x)dx +yk_y,\label{s}
\end{equation}
where
\begin{equation}\label{ph}
%p_e(x)=\sqrt{\left(\frac{E}{\hbar v_F}\right)^2-k_y^2},\\
p_h(x) = \sqrt{\left(\frac{U_0-E}{\hbar v_F}\right)^2-k_y^2-e_{xx} - \frac{\beta}{a} \chi (x)}.
\end{equation}
In the scattering process, the momentum along y-direction is conserved. Thus, we can write, $\partial_{k_y} S(k_y,x,y)=0$ and find the semiclassical trajectories of the beams as
\begin{equation}
y = -x_s\frac{k_y}{\sqrt{\left(E/\hbar v_F\right)^2-k_y^2}}-\partial_{k_y} \int_{0}^{x} p_h(x)dx.\label{y}
\end{equation}
For the strain in p-region in the vicinity of interface ($x=0$), we have $\varepsilon_{xx}=0$ and $\partial_x \varepsilon_{xx}=-Aq^2$. Thus, we can write Eq.~(\ref{y}) as~\cite{tian17}
\begin{equation}\label{y1}
y = -x_s \tan\phi -x \tan\theta,
\end{equation}
where
\begin{align}
\tan\phi &= \frac{k_y}{\sqrt{(E/\hbar v_F)^2-k_y^2}},\\ 
\tan\theta &= -\frac{k_y}{\sqrt{((U_0-E)/\hbar v_F)^2-k_y^2+\tau Aq^2}}.
\label{tantheta}
\end{align}
The strain induced magnetic field can  modify semiclassical trajectories and induce valley splitting due to the action of the valley dependent Lorentz force. The schematic diagram for valley splitting of the beams is shown in Fig.~\ref{fig3}(a), where the strain is applied to the whole p-region through the bottom gate while preserving the momentum, $p_y=\hbar k_y$, along the y-direction. The distribution of strain in the p-region is shown in Fig.~\ref{fig3}(b).  The strain engineering of such kind at the device level is experimentally feasible in graphene nanoribbons~\cite{meng13}.
Note that the particle trajectories corresponding to the Dirac points $K$ and $K'$ experience equal but opposite fictitious magnetic fields. This unique behavior leads to valley splitting of the trajectories in the Veselago lens focusing in the setup shown schematically in Fig.~\ref{fig3}(a). Simulations of particle trajectories obtained from the semiclassical action in the presence of fictitious magnetic fields are shown in Fig.~\ref{fig3}(c). The fictitious magnetic fields are shown in the   background image in Fig.~\ref{fig3}(c). As can be seen in Fig.~\ref{fig3}(c), we find the valley splitting due to the Lorentz force induced by fictitious magnetic fields.
\begin{figure*}
\includegraphics[width=17.0cm,height=3.5cm]{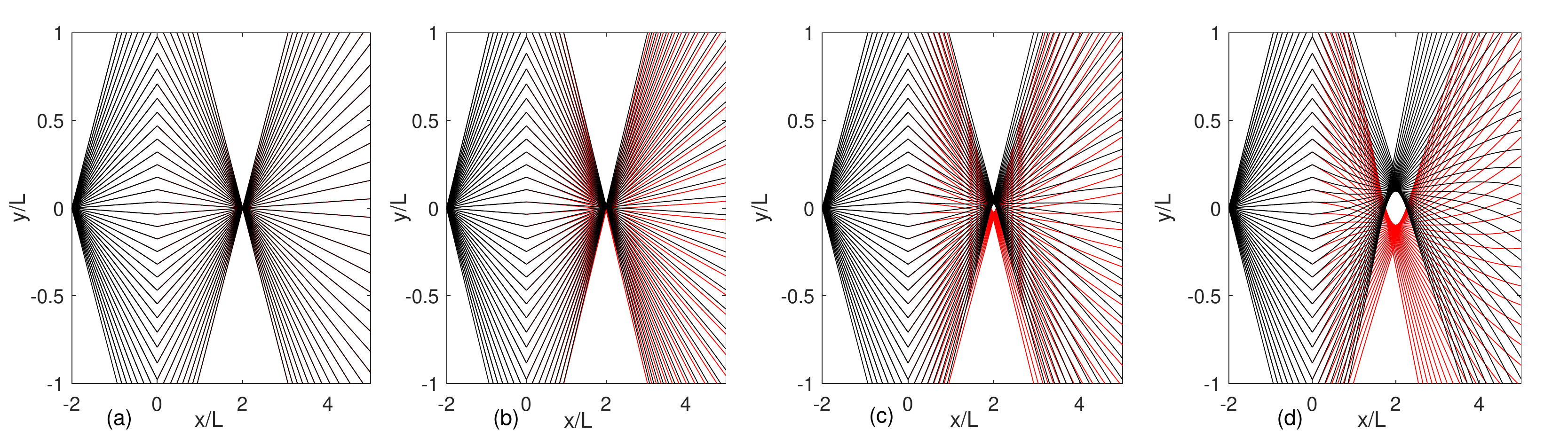}
\caption{\label{fig2splitting} Simulations of electron-hole beams trajectories for unstrained case in (a) and for strained cases (b-d). The dimensionless parameters for amplitudes are chosen as $A=-0.001$ in (b), $A=-0.005$ in (c) and $A=-0.02$ in (d). Other parameters are same as to Fig.~\ref{fig3}. For $L=100nm$, one can map these dimensionless strain amplitude in terms of fictitious magnetic fields, $Bs=0.1T, 0.6T, 2.3T$ in (b,c,d). 
}
\end{figure*}

\begin{figure*}
\includegraphics[width=18.0cm,height=3.5cm]{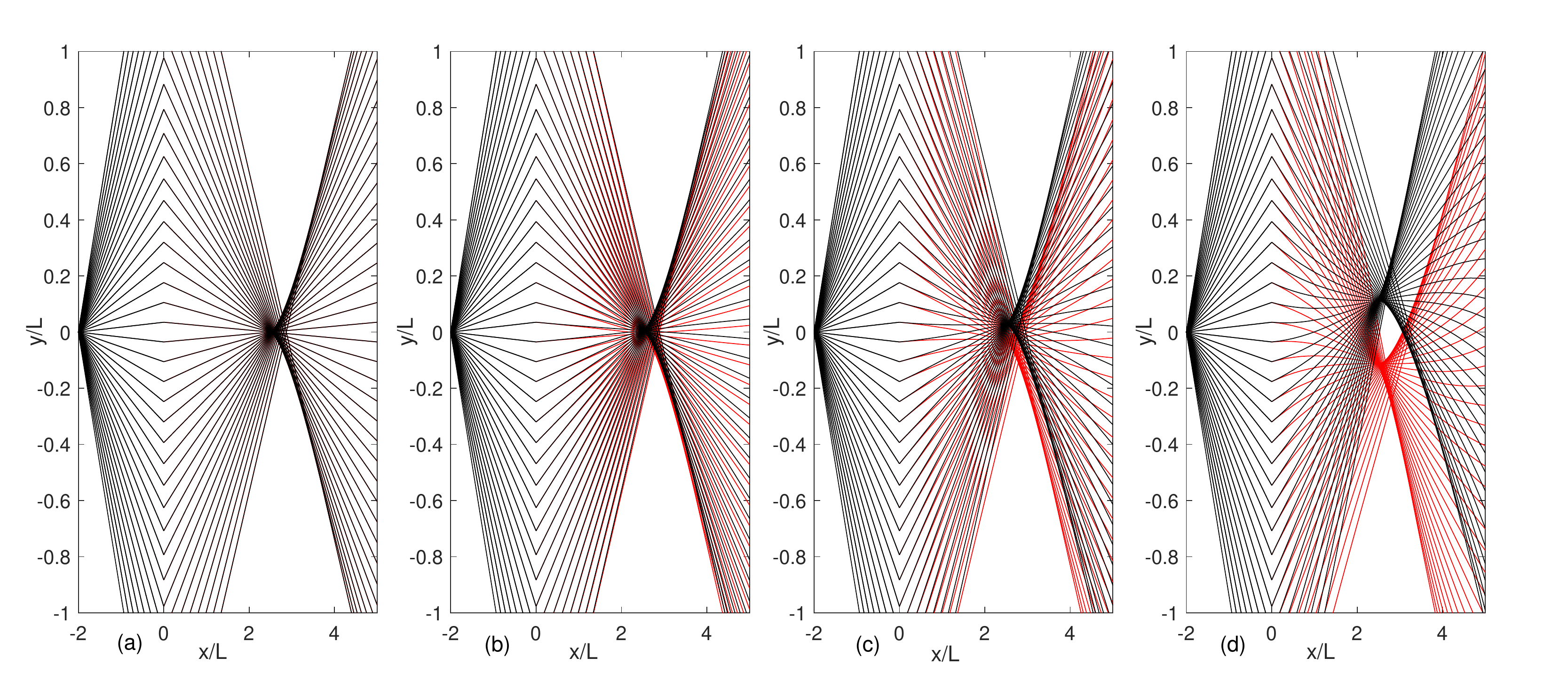}
\caption{\label{fig2splittingCr} The parameters and strain magnitude are chosen the same as in Fig.~\ref{fig2splitting} but E =2 and $U_0$=4.5. The fictitious magnetic field due to strain leads to the valley splitting and modifies the caustics pattern.
}
\end{figure*}
\begin{figure*}
\includegraphics[width=18.0cm,height=3.5cm]{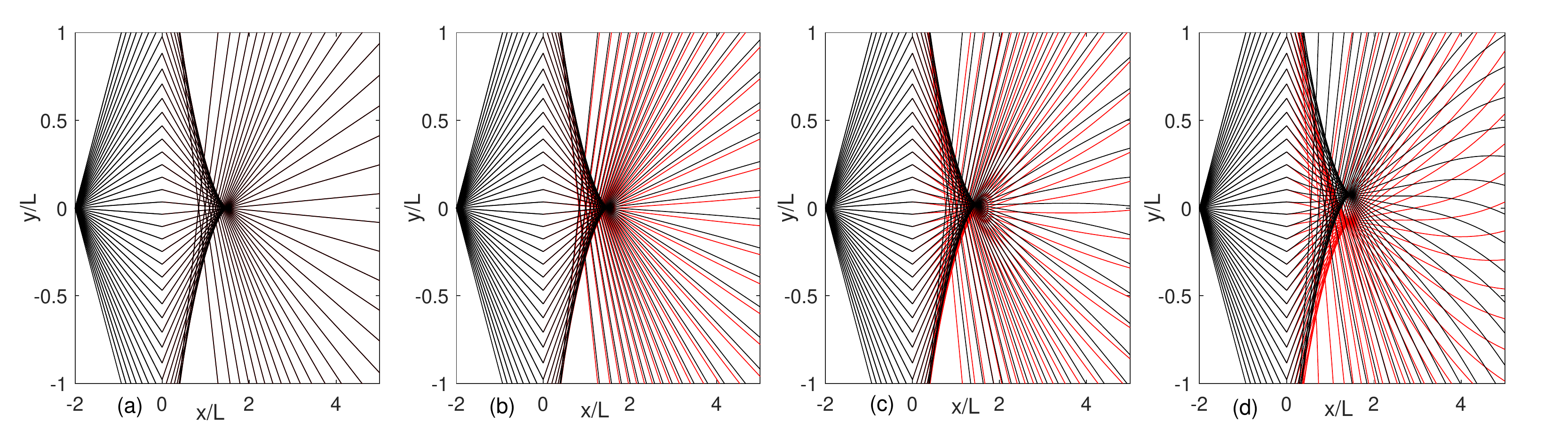}
\caption{\label{fig3splittingCl} The parameters and strain magnitude are chosen the same as in Fig.~\ref{fig2splitting} but E =2 and $U_0$=3.5. The fictitious magnetic field due to strain leads to the valley splitting and modifies the caustics pattern.
}
\end{figure*}

In Fig.~\ref{fig4Diff}, we provide the simulation results for the diffraction patterns in the p-region for unstrained (upper panel) and strained (lower panel) cases. As can be seen in Fig.~\ref{fig4Diff}(a) for unstrained symmetric case $(U_0=2E)$, a symmetric diffraction pattern, i.e., the ideal case for hole focusing, is observed. For asymmetric cases of unstrained graphene ($U_0>2E$ or $U_0<2E$), we find the caustics to the right and to the left of the focal point in Figs.~\ref{fig4Diff}(d) and \ref{fig4Diff}(g), respectively. The caustics arise when the classical trajectories have an envelope. By considering strain in the p-region for symmetric case, as shown in Fig.~\ref{fig4Diff}(b), in addition to the valley splitting, we also observe caustics. These can be also seen in Fig.~\ref{fig3}.  For the strained, asymmetric cases ($U_0>2E,$ or, $U_0<2E$) in Fig.~\ref{fig4Diff}(e,h), we again find valley splitting and somewhat deformed pattern of caustics.

In Figs.~\ref{fig2splitting}, \ref{fig2splittingCr}, and \ref{fig3splittingCl}, we investigate the evolution of the caustics for different magnitudes of strain. For symmetric case ($U_0=2E$) in Fig~\ref{fig2splitting}, we find that the caustics accompany the valley splitting in agreement with the diffraction patterns calculated earlier. For asymmetric case in Figs.~\ref{fig2splittingCr} ($U_0>2E$) and \ref{fig3splittingCl} ($U_0<2E$), we observe the deformation of caustics by the effect of strain. This deformation can be significant for larger strain ($B_s>8T$), resulting in the valley separation.

Finally, we note that the validity of semiclassical approximation requires electron wavelength, $\lambda_e=2\pi \hbar v_F/E<<\ell_B$, where $\ell_B=\sqrt (\hbar /e B_f)$ is the magnetic length induced by fictitious magnetic fields. For parameters chosen in our calculations, we estimate $\lambda_e\sim nm$ and $\ell_B\sim \mu m$ where $E$ approximately corresponds to the Fermi energy at room temperature as in most experiments~\cite{levy10}.  The system size is also chosen to be much larger than $\lambda_e$.  
\begin{figure}
\includegraphics[width=8.5cm,height=6cm]{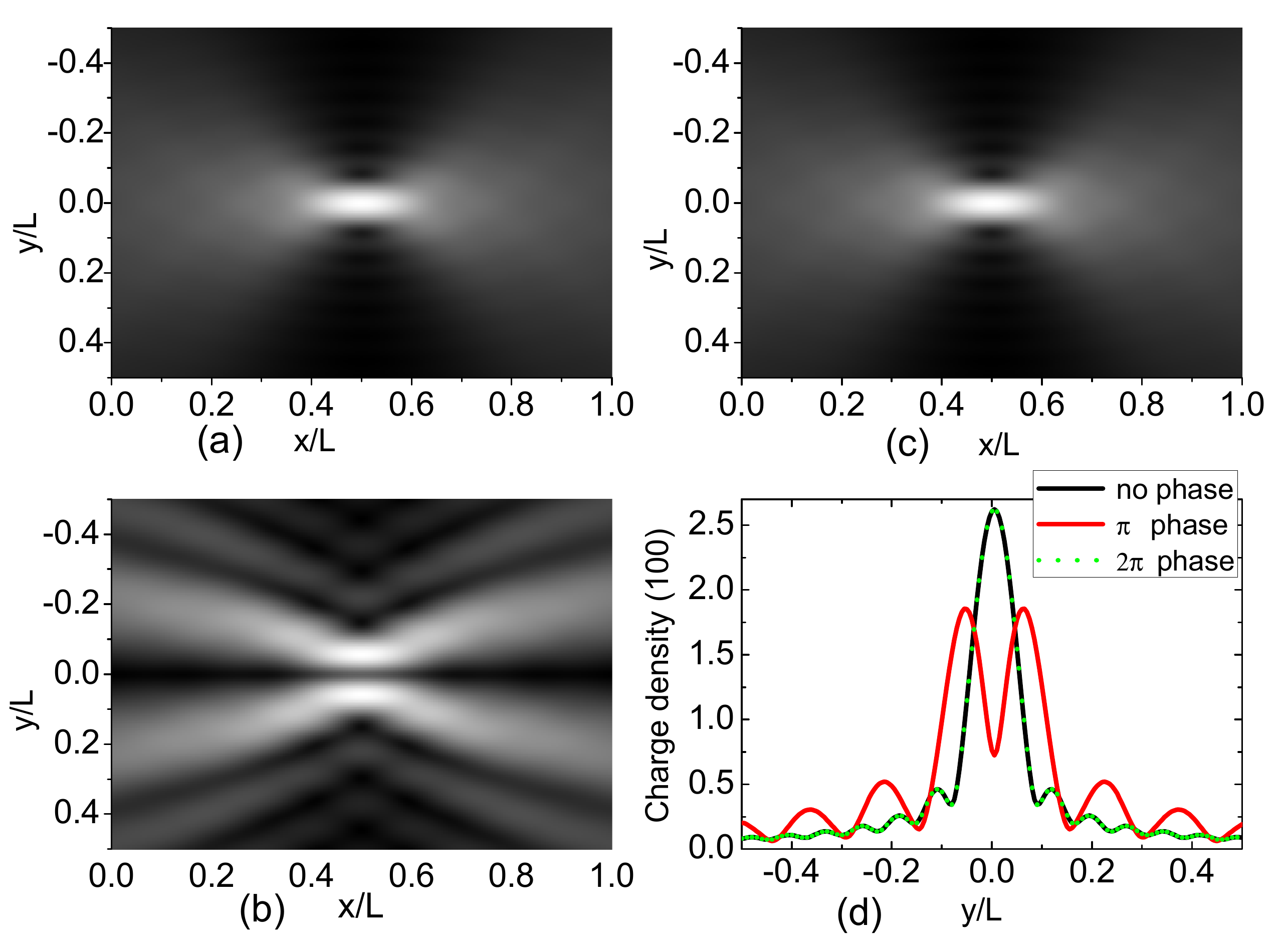}
\caption{\label{fig2AB} Diffraction patterns of the particle density for vanishing vector potentials (i.e., having no AB phase) in (a),  $\pi$ AB phase (i.e., destructive interference) in (b), and $2\pi$ AB phase (i.e., constructive interference) in (c). The cross-section plot of particle density along y-direction passing through the focal point is shown in (d), which also captures the maxima of particle density in (a), (b) and (c). The dimensionless parameters are chosen as $\iota=0.0464$ and $U=2E=4$. For the case of realistic graphene p-n junction, these dimensionless numbers correspond to $L=200$ nm and  $U_0=2E=0.28$ eV. 
}
\end{figure}

\subsection{Aharonov-Bohm phase}\label{AB}
To study signatures of strain-induced AB-like phase, we consider strain produced by an insertion shown in Fig.~\ref{strain1}, which could be a result of a dislocation~\cite{Yazyev.LouiePRB2010,bonilla12science,carpio08njp,shallcross17nc} or a line defect~\cite{Berger.RatschPRB2016,alexandre17prb,ishikawa18nc}. To account for the physics associated with AB phase, we write the components of effective in-plane displacement in polar coordinates as
\begin{eqnarray}
u_r=\frac{b_y}{2\pi}(\pi-\left|\theta\right|)\sign \theta,\label{ur}\\
u_\vartheta=\frac{b_x}{2\pi}(\pi-\left|\theta\right|)\sign \theta,\label{ut}
\end{eqnarray}
where the polar angle $\theta$ is measured from the insertion, and $\vec{b}$ describes the insertion (see Fig.~\ref{strain1}). One can notice that the vector potential corresponding to Eqs.~(\ref{ur}) and (\ref{ut}) results in an AB-like phase. When the insertion results from a dislocation in a graphene layer, we get $\vec{b}=n \vec{a}_1 +m \vec{a}_2$, where
$\vec{a}_1$ and $\vec{a}_2$ are translation vectors of graphene lattice. In what follows, we will characterize dislocation defects by two numbers as $(n,m)$. The Berry phase corresponding to a dislocation defect when $\beta=0$ then becomes~\cite{Mesaros.Sadri.eaPRB2009}
\begin{equation}
\Phi=\boldsymbol K_\pm \cdot \boldsymbol b,
\end{equation}
where we introduced the vectors $\boldsymbol K_+=\boldsymbol K'$ and $\boldsymbol K_-=\boldsymbol K$ to characterize the two valleys. 
We note here that the Berry phase flips the sign depending on the valley and only the phases given by $2\pi/3$ or $-2\pi/3$ can be accumulated on defects induced by dislocations~\cite{Mesaros.Sadri.eaPRB2009}. We argue that other non-quantized phases may be accumulated in the setup shown in Fig.~\ref{strain1} due to strain. To account for such situations,  we also consider arbitrary, non-physical values of $b$. Furthermore, we note that the strain-induced Berry phase can combine with the real AB phase due to a magnetic field flux~\cite{Mesaros.Sadri.eaPRB2009}.

We now discuss the effect of the Berry phase in the setup shown in Fig.~\ref{strain1}.
The semiclassical trajectories with $k_y>0$ accumulate $(+\Phi /2)$ AB-like phase but trajectories  with $k_y<0$ accumulate $(-\Phi/2)$ AB-like phase. When the two beams are recombined at the focal point, the total wavefunction has the form
\begin{equation}\label{Psit}
\Psi=\psi_1\exp(i\Phi/2)+\psi_2\exp(-i\Phi/2),
\end{equation}
where $\psi_{1/2}=G(x,x_s)(\alpha_1~\alpha_2)^T$ is the Green's function calculated without strain. It is clear that by tuning the phase, $\Phi$, we can tune the system between destructive and constructive interference.
This approach is well justified for trajectories away from the center region. Furthermore, for a typical system size the role of trajectories passing through the center is negligible.
We also confirm the validity of such semiclassical approximation by performing the tight binding simulations of transport in the presence of strain using Pybinding~\cite{moldovan_dean_2017} and Kwant~\cite{Groth.Wimmer.eaNJoP2014} packages.

In Fig.~\ref{fig2AB}, we plot the results of semiclassical calculations for the particle density in the vicinity of the focal point for three cases demonstrating the destructive and constructive interference: (a) zero flux, (b) $\pi$-flux and (c) $2\pi$-flux. 
Here for simplicity we assume that electrons are valley polarized.
Evidently, we find destructive interference patterns for $\pi$ AB phase,  and constructive interference patterns for $2\pi$ AB phase. The phase shifts for constructive and destructive interference patterns are also reflected in Fig.~\ref{fig2AB}(d). This shows that realization of AB phase in Veselago lens focusing by strain engineering is experimentally feasible.
Here the constructive and destructive interference patterns can be observed by tuning mechanical properties of strain in a controllable way, for example, by controlling the chemical line defects~\cite{Berger.RatschPRB2016,alexandre17prb,ishikawa18nc} and dislocations~\cite{Yazyev.LouiePRB2010,bonilla12science,carpio08njp,shallcross17nc}.

\begin{figure}
\includegraphics[width=\columnwidth]{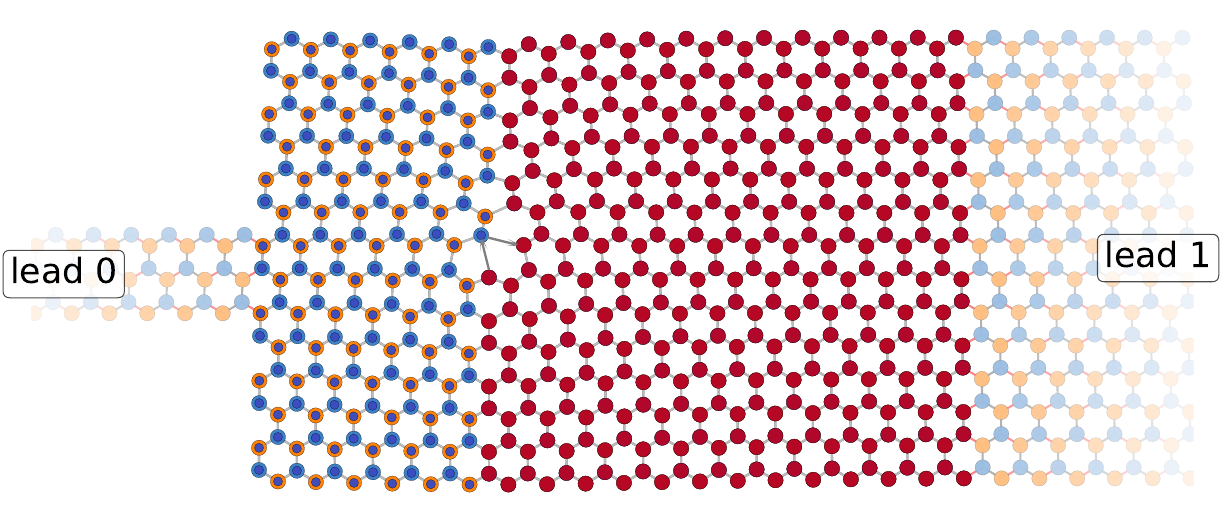}
\caption{\label{tb} Schematics of a graphene p-n junction with attached leads. The hole region is shown by dark red with a wide infinite lead while the electron region is shown by dark blue with a narrow infinite lead. A dislocation in the center of the figure corresponds to the Berry phase $\Phi=\pm 2\pi/3$.
}
\end{figure}

\begin{figure}
\includegraphics[width=\columnwidth]{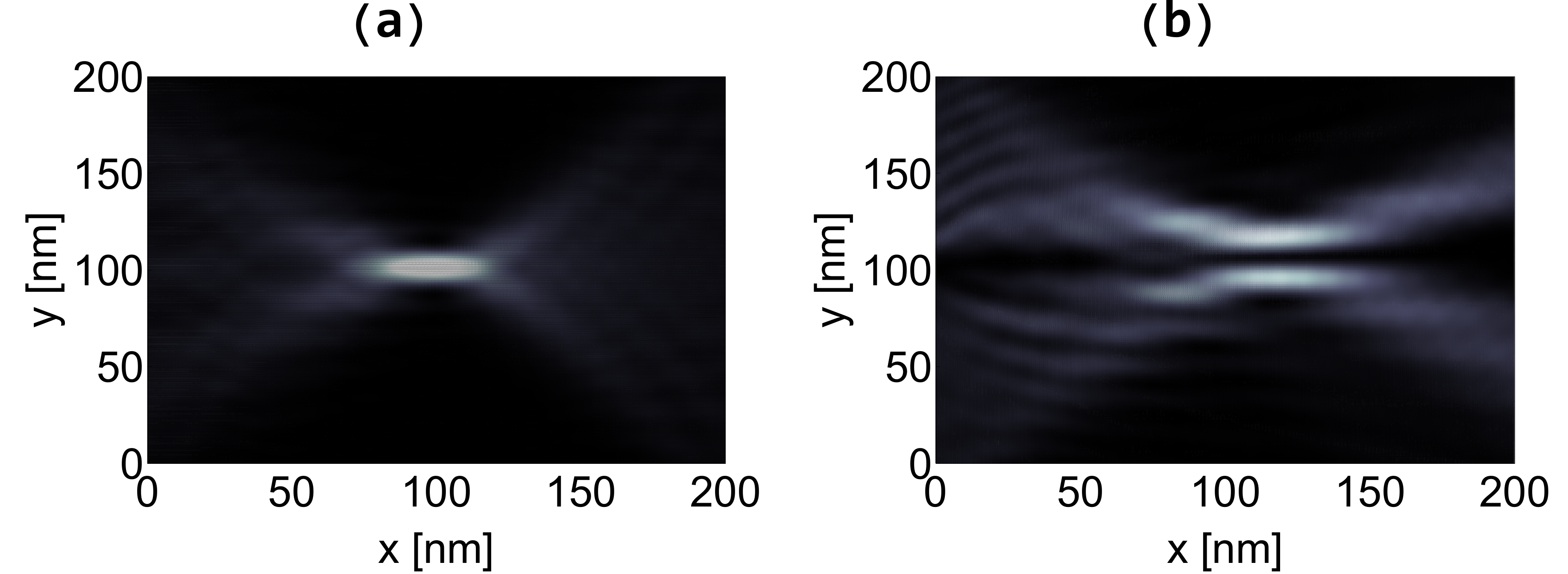}
\caption{\label{ABtb} Diffraction patterns of the particle density obtained by the tight-binding simulations for vanishing strain in (a), and for a strain corresponding to a $(2,0)$ dislocation with the Berry phase $\Phi=\pm 2\pi/3$ in (b). The plots correspond to $U_0=2E=0.2t$ where $t=$2.8 eV  is the hopping parameter.
}
\end{figure}

\subsection{Tight-binding approach}\label{tba}

We now simulate the behavior of charge carriers in a  graphene p-n junction using a lattice model of graphene in a tight-binding approach for transport calculations implemented in Kwant~\cite{Groth.Wimmer.eaNJoP2014} package. We simulate a two-terminal system schematically shown in Fig.~\ref{tb} where the injected electrons have no valley polarization. Thus, here we show that interference can be also demostrated on electrons with no valley polarization. Electrons are injected into the device through a narrow lead and then collected through a drain lead. The strain is applied to the system using tools included in a Pybinding code~\cite{moldovan_dean_2017}.

It is customary to express transport responses in terms of the retarded (advanced) Green's functions:
\begin{equation}
G^{r(a)}(E)=[E-H-\Sigma^{r(a)}]^{-1},
\end{equation}
where the tight-binding Hamiltonian only accounts for the central region while coupling to the leads is included in the self-energy $\Sigma^{r(a)}$.
At zero temperature, the conductance can be calculated from the expression;
\begin{equation}
G=\frac{e^2}{h} \Tr[\Gamma_r G^r \Gamma_l G^a],
\end{equation}
where $\Gamma_{l(r)}=i(\Sigma^r_{l(r)}-\Sigma^a_{l(r)})$ corresponds to the broadening due to the left (right) lead. For the purposes of demonstrating the Veselago lens effect, one can also consider the electron density response to electrons injected through a narrow lead. Here we calculate such response given by
\begin{equation}
\delta \rho(i)=\frac{e^2}{2 \pi} \Tr [G^r \Gamma_l G^a]_{ii},
\end{equation}
where the trace is taken over a unit cell and the index $i$ enumerates unit cells.

In Fig.~\ref{ABtb}, we plot the local particle density calculated for the strain magnitude corresponding to a dislocation (see Fig.~\ref{strain1}). The accumulated Berry phase, $\Phi=\boldsymbol  K_\pm \cdot \boldsymbol b$, combined with the strain induced phase clearly leads to interference at the focal point of the Veselago lens. In Fig.~\ref{ABtb}(b), the insertion corresponds to $(2,0)$ dislocation resulting in the Berry phase $\Phi=\pm 2\pi/3$. One can see a clear destructive interference pattern at the focal point. Finally, we also observe the recovery of the peak at the focal point for a $(3,0)$ dislocation (not shown in the figure). Our results are consistent with the semiclassical approach, as discussed in details in the previous sub-section.

\section{Conclusion}\label{conclusion}
%\noindent
%\textit{\underline{Conclusions}.}
We have shown that the presence of strain can lead to a Lorentz force and accumulation of the Berry phase in a Veselago lens based graphene p-n junction. We have shown that both effects can be separately identified. 
In particular, we have demostrated the valley separation and signatures of strong Lorentz force in the trajectories of graphene holes and electrons in corrugated graphene~\cite{meng13}, which could have implications for the field of valleytronics. The valley split current can be further measured in a setup similar to Ref.~[\onlinecite{Lee.Park.eaNP2015}]. We have also demonstrated how the same p-n junction can be used to study the Berry phase accumulation in a setup containing line defects and dislocations. This analog of the 
Aharonov-Bohm phase can be identified by observing interference patterns, and in the presence of the real magnetic field will also have additional component corresponding to the real field.
Our ideas can lead to experiments in which one can map the strain by analyzing the interference patterns in electron optics devices. In addition, the signatures of Lorentz force in the trajectories of graphene holes and electrons can have implications for the field of valleytronics.

\begin{acknowledgments}
SP was supported by the NSF MRSEC grant No. DMR-1420645, Canada Research Chair (CRC) program and Natural Sciences and Engineering Research Council (NSERC) of Canada. AAK and RN were supported by the DOE Early Career Award DE-SC0014189. RM was supported by CRC program and NSERC of Canada. The
computations were performed utilizing the Holland Computing
Center of the University of Nebraska.
\end{acknowledgments}

\bibliography{bib}

\end{document}